\documentstyle[epsf]{mn2e}

\DeclareMathAlphabet{\mathbi}{\encodingdefault}{\rmdefault}{\bfdefault}{\itdefault}
\DeclareRobustCommand{\bit}[1]{\ifmmode\mathbi{#1}\else\textbf{\textit{#1}}\fi}
\let\bolditalic=\bit

\newcommand{\be}{\begin{equation}}
\newcommand{\ee}{\end{equation}}
\def\matr#1{{\sf {\bf #1}}}   
\def\vect#1{{\bolditalic #1}}
\def\thetavec{\vect{\theta}}
\def\A{{\bf A}} 

\def\C{{\bf C}}
\def\F{{\bf F}}
\def\L{{\bf\Lambda}} 
\def\G{{\bf\Gamma}} 
\def\a{{\bf a}} 
\def\c{{\bf c}}
\def\m{{\bf m}}
\def\r{{\bf r}}
\def\x{{\bf x}} 
\def\T{{\bf \Theta}} 
\def\zero{{\bf 0}} 
\def\Aplus{{\bf A}^{^+}} 
\def\Aminus{{\bf A}^{^-}} 
\def\eq#1{equation~(\ref{#1})}

\title[WSLAP: Reconstructing W\&S lensing] 
{Combined reconstruction of weak and strong lensing data with WSLAP}  
\author[Diego et al.]  
  {J.M. Diego$^1$, M. Tegmark$^1$, P. Protopapas$^2$, H.B. Sandvik$^3$.\\  
   $^1$ MIT Kavli Institute for Astrophysics and Space Research, Cambridge, MA 02139, USA.\\
   $^2$ Harvard-Smithsonian Center for Astrophysics, 60 Garden Street, Cambridge, MA 02138, USA.\\
   $^3$ Max-Planck-Institut f\"{u}r Astrophysik, D-85748 Garching, Germany.}

\date{Draft version \today}  
  
\pagerange{\pageref{firstpage}--\pageref{lastpage}}  
  
\begin{document}  
\maketitle  
  
\label{firstpage}  
\begin{abstract}  
We describe a method to estimate the mass distribution of a gravitational lens and the 
position of the sources from combined strong and weak lensing data. 
The algorithm combines weak and strong lensing data in a unified way  
producing a solution which is valid in both the weak and  
strong lensing regimes. 
We study how the result depends on the relative weighting of the weak and strong lensing data
and on choice of basis to represent the mass distribution. 
We find that combining weak and strong lensing information has two major advantages: 
it eliminates the need for priors and/or regularization schemes for the intrinsic size of 
the background galaxies 
(this assumption was needed in previous strong lensing algorithms)
and it corrects for biases in the recovered mass in the outer regions where the strong 
lensing data is less sensitive.
The code is implemented into a software package called WSLAP (Weak \& Strong Lensing 
Analysis Package) which is publicly available at http://darwin.cfa.harvard.edu/SLAP/.

\end{abstract}  

\section{Introduction}\label{section_introduction}  
Lensing problems usually distinguish between two regimes, strong and weak. 
In the strong lensing regime, a background source galaxy appears as multiple images,
while in the weak lensing regime, its image suffers a small distortion 
which typically elongates it in a 
direction orthogonal to the gradient of the potential. 
The two problems are normally 
studied separately and, at best, they are combined afterward. Only a few attempts have 
been made to combine both regimes in the same analysis 
(e.g. Bradac et al. 2005, Broadhurst, Takada, Umetsu et al. 2005b).\\

The quality and quantity of strong and weak lensing data is growing 
rapidly, motivating the use of algorithms capable of making
full use of the amount of information present in the images. In the
early years of strong lensing data analysis, it was common to have only few constraints 
to work with. The small number of constraints made it impossible to
extract useful information about the mass distribution of the
lens without invoking a simple parametrization of the lens or the
gravitational potential (Kneib et al. 1993, 1995, 1996, Broadhurst et
al 1995, 2005a, Sand et al. 2002, Gavazzi et al. 2004). 
The common use of parametric models requires making educated guesses  
about the cluster mass distribution, for instance that the dark matter 
halos trace the luminosity of the cluster or that galaxy profiles 
possess certain symmetries.\\

Nowadays, it is possible to obtain strong lensing images around the center of  
galaxy clusters with hundreds of arcs (Broadhurst et al. 2005a), where each arc contributes 
with several effective constraints in the process of solving for the
projected mass distribution of the lens.  
In addition, weak lensing measurements provide shear
constraints over a larger field  
of view. When added together, the number of constraints can
be sufficiently high that 
non-parametric methods can be used in the reconstruction 
of the mass. 
With such a large number of constraints, non-parametric 
methods have a chance to compete 
with the parametric ones, complementing their results and raising 
interesting questions if significant disagreements are found between
the two methodologies. 

Non-parametric approaches have been previously explored in
several papers 
(Saha et al. 1997, Abdelsalam et al. 1988b, 1998c, Trotter et al. 2000, Williams \&
Saha 2001, Warren \& Dye 2003, Saha \& Williams 2004, Bradac et al. 2005,
Treu \& Koopmas 2004) and more recently in Diego et al. (2005a) 
(hereafter {\bf paper I}). 
In paper I, the authors showed that it is possible to
non-parametrically reconstruct a generic mass profile (with
substructure) provided that the number of strongly lensed arcs with known redshifts is
sufficiently large. They developed a  package called SLAP upon which WSLAP is based.
Paper I also showed how working with extended images rather than
just their positions adds enough constraints to solve 
the regularization problem found 
in other non-parametric algorithms  (see Kochanek et al. 2004 for a
discussion of this issue).  An application of the method using SLAP
on data from A1689 can be found in Diego et al. 2005b 
(hereafter {\bf paper II}).

Most of the literature on lensing observations is based
on either weak or strong  
lensing data. Only few papers have attempted to combine both weak and
strong lensing data 
(e.g Bradac et al. 2005, Broadhurst, Takada, Umetsu et al. 2005b).
The main advantage of combining both regimes is that they complement
each other, filling the gaps  
and correcting the deficits of each other. Strong lensing data is
particularly sensitive to the central  
mass distribution of the lens but is relatively insensitive to the
outer regions. On the other hand, weak  
lensing cannot capture the fine details in the central regions but
can trace the mass distribution  
further out than strong lensing data. One of the problems of modeling weak
lensing data is the so-called mass-sheet  
degeneracy. Strong lensing can break this degeneracy if several arcs
are observed and the sources of  
these arcs span a wide range of redshifts. 
Some algorithms have been proposed for combining the weak and strong 
lensing regimes 
(Abdelsalam et al. 1998a; Bridle et al. 1998;
Saha, Williams and Abdelsalam 1999;
Kneib et al. 2003;
Smith et al. 2004; 
Bradac et al 2004), 
but usually they need to assume a prior on the mass (Kneib) or luminosity (Abdelsalam) 
or regularize the problem (Bradac, Abdelsalam). 
One of the purposes of this paper is to show how the above assumptions
can be eliminated and that a well defined likelihood can be defined
for the combined weak and strong lensing data set. 

\section{WSLAP and strong lensing}
The fundamental problem in lens modeling is the following: Given the   
$N_{\thetavec}$ positions of lensed images, $\matr{\thetavec}$, what are the corresponding positions $\matr{\beta}$ of the 
background galaxies and the mass distribution $m(\thetavec)$ of  
the lens? Mathematically this entails inverting the lens equation  
\begin{equation}  
\vect{\beta} = \thetavec - \vect{\alpha}(\thetavec,m(\thetavec))  
\label{eq_lens}  
\end{equation}  
where $\vect{\alpha}(\thetavec)$ is the deflection angle created by the lens   
which depends on the observed positions, $\thetavec$. Each
observed position $\thetavec$  
contributes with two constraints, $\thetavec=(\theta_x,\theta_y)$, so we have 
$2N_{\thetavec}$ strong lensing constraints. 

The deflection angle $\alpha$ at the position $\thetavec$, is found by  
integrating the contributions from the whole mass distribution:  
\begin{equation}  
\alpha(\theta) = \frac{4G}{c^2}\frac{D_{ls}}{D_s D_l} \int m(\theta')  
                 \frac{\theta - \theta'}{|\theta - \theta'|^2} d\theta'  
\label{eq_alpha}  
\end{equation}  
where $D_{ls}$, $D_l$, and $D_s$ are the angular distances  
from the lens to the source galaxy, the distance from the observer to  
the lens and the distance from the observer to the source galaxy  
respectively. In equation    
(\ref{eq_alpha}) we have made the usual thin lens approximation   
so the mass $m(\theta')$ is the projected mass along the line of   
sight  $\theta'$. 
From the deflection angle one can easily derive the magnification produced 
by the lens at a given position:
\begin{equation}
\mu^{-1}(\thetavec) = 1 - \frac{\partial\alpha_x}{\partial x} - \frac{\partial\alpha_y}{\partial y} 
                          + \frac{\partial\alpha_x}{\partial x}\frac{\partial\alpha_y}{\partial y} 
                         - \frac{\partial\alpha_x}{\partial y}\frac{\partial\alpha_y}{\partial y} 
\end{equation}

We find it convenient to expand the projected mass distribution
in an set of basis functions:
\begin{equation}
m(x,y) = \sum_l c_l f_l(x,y),
\label{eq_F}
\end{equation}
where $f_l(x,y)$ are the basis functions and $c_l$ the
coefficients of the decomposition.  
Here $f_l(x,y)$, can be any sort of 2D function. For instance,
one can choose orthogonal  
polynomials like the Legendre or Hermite polynomials. Or one can use
Fourier or wavelet functions as  
the basis. We find that the best results are obtained
using compact basis functions defined on a gridded version of the mass distribution 
like the ones used in papers I and II, since using extended ones tends to
over-produce arcs in the final result --- see however Sandvik
et. al. (in preparation) for a novel approach to this problem. 
In papers I and II we used for $f_l$ Gaussians with
varying widths defined in  
a multi-resolution grid. In this paper we will focus on compact bases and 
will compare the results using three different compact bases. 

After decomposing the mass as in equation (\ref{eq_F}), equation 
(\ref{eq_alpha}) can be rewritten as
\begin{equation}  
\alpha(\theta_j)= \lambda_j \sum_l c_l \int f_l(\theta')  
                 \frac{\theta - \theta'}{|\theta - \theta'|^2} d\theta' 
                =  \lambda_j \sum_l c_l\tilde{f}_l(\theta_j),
\label{eq_alpha2}  
\end{equation}  
where all the constants and distance factors are absorbed into the variable 
$\lambda_j$. Note that there is a different $\lambda_j$ for each source since 
$\lambda_j$ includes the distance factors $D_l$, $D_s$ and $D_{ls}$ which vary for 
each source.
The factor $\tilde{f}_l(\theta_j)$ is the convolution of the basis function $f_l$ 
with the kernel $(\theta - \theta')/|\theta - \theta'|^2$ evaluated at the 
point $\theta$:
\begin{equation}
\tilde{f}_l(\theta_j) \equiv \int f_l(\theta')  
                 \frac{\theta - \theta'}{|\theta - \theta'|^2} d\theta'.
\end{equation} 
If now we define the matrix $\matr{\Upsilon}$ by 
\begin{equation}
\Upsilon_{jl} = \lambda_j\tilde{f}_l(\theta_j),
\label{eq_Upsilon_ij}
\end{equation}
then all the constraints given by equation (\ref{eq_lens}) can be expressed in the simple form 
\begin{equation}
\T = \matr{\Upsilon}\c - \vect{\beta}.
\label{eq_lens2}
\end{equation}
where $\T$ is the array (vector) containing all the $\thetavec$ positions 
($\theta_x$ and $\theta_y$). 
The matrix $\matr{\Upsilon}$ has a straightforward physical interpretation: 
the element $\Upsilon_{jl}$ is just the deflection 
angle created by the basis function $f_l$ at sky position $\theta_j$.
Note that since $\theta_j$ has two components (the $x$ and $y$ components),
the are two corresponding elements in $\matr{\Upsilon}$. 

If we group all the unknowns in our problem (both $\beta$ and $\c$ or the mass in each cell) 
into a new vector $\x$, then equation (\ref{eq_lens2}) can be rewritten in the more 
compact form
\begin{equation}
\T = \L \x,
\label{eq_lens3}
\end{equation}
where $\L$ is a $2N_{\theta}\times (N_c+2N_s)$-dimensional matrix 
and $\x$ is the $(N_c+2N_s)$-dimensional vector containing all the 
unknowns in our problem (see paper I), {\it i.e.}, the $N_c$ cell masses $m_l$ (or coefficients $c_l$), 
and the $2N_s$ central positions $\beta_o$ ($x$ and $y$) of the $N_s$ sources.

\section{Adding weak lensing}
So far we have focused on solving a system of linear equations corresponding 
to strong lensing data. 
If weak lensing information is available, it can be easily incorporated 
into equation (\ref{eq_lens3}), allowing us to find the combined solution of the 
weak plus strong lensing as we will see below.


Given the gravitational potential $\psi$,  
the shear is defined in terms of the second partial derivatives of the potential 
$\psi$ (the Hessian of $\psi$):
\begin{eqnarray}
\psi_{ij}		&=&\frac{\partial^2\psi}{\partial\theta_i\partial\theta_j},\\
\gamma_1(\thetavec) 	&=& \frac{1}{2}(\psi_{11} - \psi_{22})
                        = \gamma(\thetavec)\cos[2\varphi],\\
\gamma_2(\thetavec) 	&=& \psi_{12} = \psi_{21} =\gamma(\thetavec)\sin[2\varphi],
\end{eqnarray}
where $\gamma(\thetavec)$ is the amplitude of the shear and $\varphi$ its 
orientation.
The shear can be computed in a way very similar to the magnification $\mu$ yielding
\begin{equation}
\gamma_1 = \frac{1}{2} \left ( \frac{\partial\alpha_x}{\partial x} -
\frac{\partial\alpha_y}{\partial y} \right ),
\label{eq_gamma1}
\end{equation}
\begin{equation}
\gamma_2 = \frac{\partial\alpha_x}{\partial y} = \frac{\partial\alpha_y}{\partial x}.
\label{eq_gamma2}
\end{equation}
The amplitude and orientation of the shear are given by 
\begin{equation}
\gamma = \sqrt{\gamma_1^2 + \gamma_2^2},
\end{equation}
\begin{equation}
\varphi = \frac{1}{2} \mathrm{atan}(\frac{\gamma_2}{\gamma_1}).
\end{equation}

Given a number of shear measurements, an equation similar to (\ref{eq_lens2}) can be 
written for the shear:
\begin{equation}
\left ( \begin{array}{c} \vect{\gamma_1} \\
\vect{\gamma_2} \end{array} \right ) = \left ( \begin{array}{c} \matr{\Delta_1} \\
\matr{\Delta_2}\end{array} \right )  \c,
\label{eq_shear2}
\end{equation}
where each element in the matrices $\matr{\Delta_1}$ and $\matr{\Delta_2}$ represents the 
contribution to  the shear ($\vect{\gamma_1}$ and $\vect{\gamma_2}$ respectively) of each 
one of the basis functions. The expression for $\Delta_{ij}$ can be easily 
derived from equations (\ref{eq_Upsilon_ij}), (\ref{eq_gamma1}) and (\ref{eq_gamma2}). 
The explicit form of $\Delta_{ij}$ is given in Diego et al. (2005d).\\

After combining the strong and weak lensing regimes by regrouping the observed 
$\thetavec$-positions of the strongly lensed galaxies and the measured shear, the  
new measurement vector $\vect{\Phi}$ will have the structure
\begin{equation}
\vect{\Phi}^t = (\theta_x,\theta_y,\gamma_1,\gamma_2), 
\end{equation}
and the corresponding system of linear equations representing the lens equation reads
\begin{equation}
 \left ( \begin{array}{c} 
 \theta_x \\
 \theta_y \\
 \gamma_1 \\
 \gamma_2 \\ \end{array} \right ) = \left ( \begin{array}{ccc}
               \matr{\Upsilon_x} & \matr{I_x} & \matr{0} \\
               \matr{\Upsilon_y} & \matr{0} & \matr{I_y} \\
 	       \matr{\Delta_1} & \matr{0} & \matr{0} \\
	       \matr{\Delta_2} & \matr{0} & \matr{0} \\
	       \end{array} \right ) 
               \left ( \begin{array}{c}
	        \c \\	 
	        \beta_x \\	 
	        \beta_y \\	 
	       \end{array} \right ),
\end{equation}
where we have explicitly expanded the matrix $\L$ and the 
vector of unknowns $\x$ into their components. In the above equation, 
the matrix $\matr{0}$ contains all zeros while the $ij$ elements in matrix 
$\matr{I_x}$ are ones if the $\theta_i$ pixel ($x$-coordinate) 
is coming from the $\beta_j$ source ($y$-coordinate) and zero otherwise. 
The matrix $\matr{I_y}$ is defined in an analogous way for the $y$-coordinates.
The above equation written in compact form is simply
\begin{equation}
\vect{\Phi} = \G\x.
\label{eq_LensMain}
\end{equation}
In summary, we have formulated the full weak and strong lensing problem in a manner were the observables 
$\vect{\Phi}$ depend linearly on the unknowns $\x$, so all the complicated physics and geometry
is conveniently encoded into the known matrix $\G$.

In principle, an exact solution for $\x$ exists if the inverse of ${\G}$ 
exists (i.e $\x =  \G^{-1}\Phi$). However, in most cases, 
${\G}$ is singular and therefore does not have an inverse (some of the eigenvalues 
are basically zero within rounding errors), so a direct 
inversion of the problem is not possible. Furthermore, even when the inverse of 
${\G}$ exists, we may not be interested in finding the exact solution,
but rather in an approximate solution of equation (\ref{eq_LensMain}). 
The reason is twofold. The definition  
of $\x$ assumed that the source galaxies responsible for the strong lensing arcs are 
point-like (that is, each source is defined only by its coordinates,
$\beta_x$ and $\beta_y$).  
This assumption is inaccurate as the galaxies will have some spatial extent, so we want 
the solution to allow for some residual in equation (\ref{eq_LensMain}). Second, for the 
mass we have assumed that it is a superposition of certain basis functions, 
say cells. 
This assumption, although a good approximation, is also partially inaccurate,
so we want to incorporate this in our analysis by allowing some residual ($|\r|>0$) in 
the lens equation. This residual is defined as
\begin{equation}
\r\equiv\vect{\Phi} - \G\x.
\label{eq_Residual}
\end{equation}

\section{Solving the lens problem}
The fundamental task we are faced with is to obtain the coefficients $\c$, describing 
the lens surface mass density, and the positions $\vect{\beta}$ of the background galaxies  
in order for their combination to explain the observed arcs $\theta$ and the shear $\gamma$. 
In the previous section, we have shown how the unknowns of the problem can be 
combined into a vector $\x$, the observed data into another vector, $\Phi$, and 
the connection between the two is given by the matrix ${\G}$. These 
three elements relate to each other through the system of linear equations 
(\ref{eq_LensMain}). 

We adopt a Bayesian approach to solving the problem, finding the solution $\x$ that maximizes the 
likelihood function.
\begin{equation}
{\cal L}(\x) = e^{-\frac{1}{2}\chi^2},
\end{equation}
where we have assumed that the residual $\r$ is Gaussian distributed.
The $\chi^2$ is defined as
\begin{equation}\label{eq_chi2}
\chi^2 = \r^t\C^{-1}\r,
\end{equation}
where $\C$ is the covariance matrix of the residual $\r$. 
We model the residuals as uncorrelated ($\C$ is diagonal) and that
the elements on the diagonal  
are equal to either $\sigma_{\theta}^2$ or $\sigma_{\gamma}^2$, where the former is associated with the expected 
residual variance in the strong lensing data and the latter is for the expected residual variance in the shear 
measurements.

We will extensively discus how to best choose $\C$ below in Section~\ref{CchoiceSec}.
For the main calculations in this paper, 
we assume that the rms error $\sigma_{\theta}$ is of the order of a few pixels in the source 
plane, and model $\sigma_{\gamma}$ as uniform over the field of view, 
equal to $0.005$ for both the $\gamma_1$ and $\gamma_2$ components. 
Errors in shear measurements can be in the range of a few percent for well 
calibrated experiments (e.g Hirata et al. 2005, Heymans et al. 2005). A value 
of $\sigma_{\gamma} = 0.005$  is at the percent level for a typical cluster.   
The covariance matrix $\C$ can also incorporate a measure of
the noise in the data both in the strong and weak lensing.  

We will now explore two alternative approaches for finding the solution that
maximizes the likelihood (minimizes the $\chi^2$).

\subsection{Bi-conjugate Gradient Method}
Substituting \eq{eq_Residual} into \eq{eq_chi2},
\begin{eqnarray}
\label{eq_R2}
\chi^2 & = &(\vect{\Phi} - \G \x)^t\C^{-1}(\vect{\Phi} - \G\x)  \\ \nonumber 
       & = & \vect{\Phi}^t\C^{-1}\vect{\Phi} -2\vect{\Phi}^t\C^{-1}\G \x +  
\x^t\G^t\C^{-1}\G \x\\ \nonumber
    & = &   b - \a^t\x + \frac{1}{2}\x^t\matr{A}\x,
\end{eqnarray}
where we have defined the constant $b\equiv\vect{\Phi}^t\C^{-1}\vect{\Phi}$,
the vector $\a\equiv 2\G^t\C^{-1}\vect{\Phi}$ and
the matrix $\A\equiv 2\G^t\C^{-1}\G$.

\begin{figure*}  
   \begin{flushleft}
   \epsfysize=9.cm   
   \begin{minipage}{\epsfysize}\epsffile{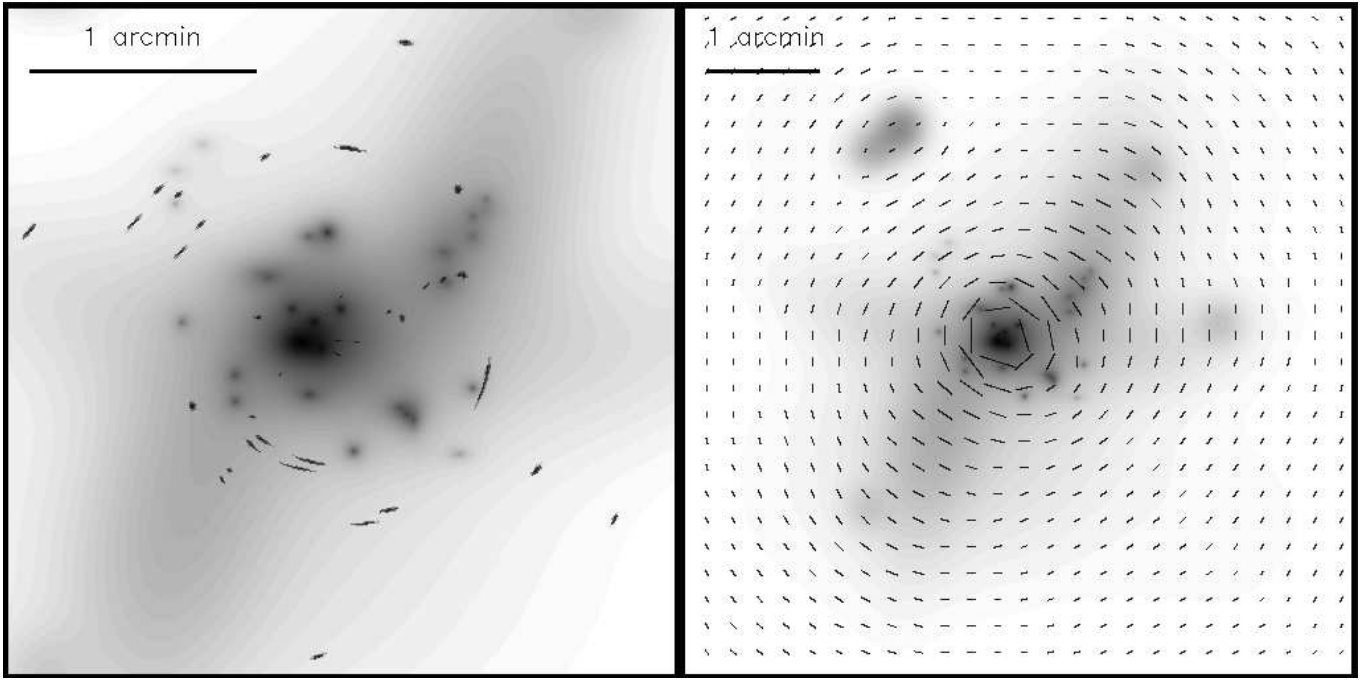}\end{minipage}  
   \caption{  
            Original mass used to test the algorithm (at z=0.4). 
            The mass is built out of a superposition of 30 NFW profiles with added 
            ellipticities. The total mass in the field of view (6 arcmin) is 
            $1.17397 \times 10^{15} h^{-1} M_{\odot}$. 
            The left panel shows the central region (3 arcmin, $0.663 \times 10^{15} h^{-1} M_{\odot}$) 
            and the strongly lensed galaxies. These arcs are lensed images of 9 galaxies between redshifts 
            1 and 6.5. The right panel (6 arcmin) shows the shear field and the outer regions of the 
           cluster.
           }  
   \label{fig_mass}
   \end{flushleft}
\end{figure*}  

Minimizing this by setting the derivative with respect to $\x$ equal to zero gives 
a formal solution $\x=\A^{-1}\a$. This is not useful in practice, however, since 
$\G$ (and therefore also $\A$) is normally rather singular.
In paper I, we found a simple regularization technique that gives physically reasonable results:
minimizing $\chi^2$ using the iterative bi-conjugate gradient method (Press et al. 1997),
but stopping once an approximate solution of equation \eq{eq_LensMain} had been found rather than 
continuing to iterate toward the formal solution.
Specifically, the bi-conjugate 
gradient method performs successive minimizations which are carried out in a series 
of orthogonal conjugate directions with respect to the metric $\A$. 
The algorithm starts with an initial guess for the solution (for instance $\x_0=\zero$). 
Then the algorithm chooses as a first minimization direction the gradient   
$\nabla\chi^2$ at $\x_0$. 
Then  it minimizes in directions which are conjugate to the previous ones until   
it reaches the minimum or the $\chi^2$ is smaller than certain target value $\epsilon$. 
We will discuss later how to choose $\epsilon$ --- we will find that 
combining weak and strong lensing makes the choice of $\epsilon$ much less relevant 
than when only strong lensing data is used in the analysis.

\subsection{Nonnegative quadratic programming}

Although the bi-conjugate gradient method is a fast and effective way to find an approximate solution,
it is not ideal.
The regularization procedure was required because certain modes in the mass distribution corresponded
to eigenvalues near zero in the matrix $\A$. Plotting these unconstrained modes shows that
they all oscillate, trading off positive mass in some places against unphysical negative mass 
elsewhere. Without regularization, the solution can include such modes of significant amplitude,
involving negative mass in certain cells.
Both the regularization problem and negative mass problem can therefore be eliminated in one
fell swoop by using a constrained minimization algorithms that only minimizes $\chi^2$
in the physically meaningful region of the parameter space $\c$ where all masses are non-negative.
Our case is particularly simple: we are minimizing a quadratic function of $\x$ 
subject to constraints on $\x$ that are linear. 
This is a well studied problem in optimization theory,
and several methods have been proposed in the context of quadratic
programming (QADP).   
In this paper we will explore the approach of Sha, Saul \& Lee (2002) known as 
multiplicative updates for nonnegative quadratic programming. 

Following Sha et al. (2002), we can minimize the quadratic objective function
\begin{equation}
f(\x) = \frac{1}{2}\x^t\A\x +\a^t\x
\label{eq_fv}
\end{equation}
subject to a non-negative mass constraint, {\it i.e.}, the constraint that $m_i \geq 0$ for all $i$, where $m_i$ is 
the mass at position $i$. Note that when compared with equation (\ref{eq_R2}), we have changed the sign of vector 
$\a$ to keep the same notation of Sha et al. (2002). 
In our vector $\x$, the mass distribution is represented by expansion coefficients $\c$ rather than cell masses $\m$,
and \eq{eq_F} shows that these two vectors are related by
\begin{equation}
\m = \F\c,
\label{eq_mx}
\end{equation}
where the element $\F_{il}$ is the value of the basis function $f_l$ at position $i$.
We can therefore make the substitution $\x = \F^{-1}\m$ in equation (\ref{eq_fv}) and 
rewrite it as a function of a transformed vector denoted $\x'$ which equals $\x$ except that 
the elements defining the mass distribution are $\m=\F\c$ rather than $\c$:
\begin{equation}
f'(\x') =  \frac{1}{2}\vect{x'}^t\A'\vect{x'} + \a'^t\vect{x'},
\label{eq_fv2}
\end{equation}
where $\a'$ is the same as $\a$ but with the elements related to masses multiplied by $\F^{-1}$, 
and $\A'$ is the same as $\A$ but with the sub-matrix related to masses multiplied by $\F^{-1}$ both from the
left and from the right.
In general, the dimensionality of 
$\x'$ can be different than the dimensionality of $\x$. However, to keep the problem
simple, we assume that the masses in equation (\ref{eq_mx}) are evaluated only at the central position of 
each cell. That makes the dimensions of $\x$ and $\vect{x'}$ equal and $m_i$ can be interpreted
as simply the total projected mass in the $i^{\rm th}$-cell.  
Since the positions $\vect{\beta}$ can be also made positive (by defining the origin of the coordinates in 
the left bottom corner of the field of view), all components in the vector $\vect{\x'}$ 
($m_i$ and $\beta_j$) have to be positive.  

In conclusion, we wish to minimize \eq{eq_fv2} subject to the constraints that all elements $x_i'\ge 0$.
We solve this problem iteratively using the multiplicative 
update technique of Sha, Saul \& Lee (2002).
For simplicity of notation, we suppress all primes from \eq{eq_fv2} below.
Let us split the matrix $\A$ into its positive and negative parts $\Aplus$ and $\Aminus$ 
such that $\A = \Aplus - \Aminus$, where $A_{ij}^+\equiv A_{ij}$ 
if $A_{ij} > 0$ and 0 otherwise and  $A_{ij}^- \equiv -A_{ij}$ if $A_{ij} < 0$ and 0 otherwise. 
The solution is iteratively updated by the rule
\begin{equation}
x_{i+1} = x_i\delta_i,
\label{eq_vi}
\end{equation}
where the multiplicative updates $\delta_i$ are defined by
\begin{equation}
\delta_i = \frac{-a_i + \sqrt{a_i^2 + 4(\Aplus\x)_i(\Aminus\x)_i}}{2(\Aplus\x)_i}.
\label{eq_vi2}
\end{equation}

It is easy to see that generic quadratic programming problems have a single unique minimum.
Let $\x^*$ denote this global minimum of $f(\x)$ (within the non-negative mass part of parameter space).
Let us prove that convergence of the iteration \eq{eq_vi2}
corresponds to this minimum $\x^*$.
At this point, one of two conditions must apply for each component $x_i^*$:
Either (i)  $x_i^* > 0$ and ${\partial f\over\partial x_i}(\x^*) = 0$ or 
(ii) $x_i^* = 0$ and ${\partial f\over\partial x_i}(\x^*)\geq 0$. 
Now since
\begin{equation}
{\partial f\over\partial x_i}(\x^*) = (\A^+\x)_i - (\A^-\x)_i + a_i,
\end{equation}
the multiplicative updates in both cases (i) and (ii) take the value 
$\delta_i = 1$, the minimum is a fixed point.
Conversely, a fixed point of the iteration must be the minimum $\x^*$.

\section{Simulations}
Testing the algorithm with simulations is essential, not only to prove 
its feasibility but also to identify its failures and weaknesses. In this paper, 
we will show some results using a simulated cluster with a particularly rich 
structure. The motivation for this is twofold. First, using a highly asymmetric 
distribution motivates the use of non-parametric methods where no assumptions about 
the distribution of the mass are needed. Second, asymmetries may play a role introducing 
biases in the result which we may want to study. \\
Also, with simulations we can test how different choices for $f_l$ and
$\C$ affect the  
result. This last step is important since $f_l$ and $\C$ are
basically the only assumptions  
made in the process of fitting the data.\\

The simulated data are made of a combination of 3 basic ingredients: \\
1) the lens mass distribution that we will try to recover,\\
2) the arcs observed in the central region of our field of view  
   that will constitute the strong lensing part of the data, \\ 
3) the shear measured over the entire field of view which will constitute the
   weak lensing part of the data.\\ 
  
\subsection{Mass distribution}
In order to test the algorithm, we will use a simulated cluster with abundant internal 
structure. The cluster is placed at redshift z=0.4. It has a highly elliptical extended 
large scale component at large scale and the central region has several clumps
surrounding the central peak.  
These clumps are generated from NFW profiles with added
ellipticities. There is also a filamentary  
component crossing the field of view. The simulated cluster has a total 
projected mass of $1.174 \times 10^{15} h^{-1} M_{\odot}$ over the field of view (6 arcmin) 
and is shown in figure \ref{fig_mass}. This field of view corresponds
to a scale of 1.35 $h^{-1}$ Mpc and normally covers more than half the virial 
radius expected for clusters with this mass.\\ 

\subsection{Strong lensing data}
To generate the arcs, we place several sources behind the cluster.
The sources have redshifts between $z=1.0$ and $z=6.5$. We consider 9
sources in this redshift  
range. The arcs produced by the combination lens-sources are shown in figure 
\ref{fig_mass} (left panel). These arcs will constitute the strong lensing part of our 
data set. We use all the pixels containing part of one arc in the previous image. There 
are 621 of these pixels. All the sources have at least two lensed
images in the previous plot.  
Some sources appear as many as five times. Although we search for
multiple images only in the central  
part of the field of view ($3\times 3$ arcmin$^2$), we use the mass over the entire field 
of view  ($6\times 6$ arcmin$^2$) to calculate the deflection angle.\\

\subsection{Weak lensing data}
For the weak lensing part, we calculate the shear field over the
entire  $6\times 6$ arcmin$^2$  
field of view (or equivalently $1.35\times 1.35$ (h$^{-1}$ Mpc)$^2$). 
The shear is simulated assuming the sources have a medium redshift of $z=3$. 
We consider that the observation is deep enough and that a shear measurement can be 
obtained after averaging areas of $14.5 \times 14.5$ 
arcsec$^2$ which renders 625 shear measurements over the field of view  
(625 $\gamma_1$ and 625 $\gamma_2$). 
The shear field is shown in the right panel of figure
\ref{fig_mass}. \\

Summarizing, the strong lensing data consist of $N_{\theta}=621$ pixels distributed in about 40 strongly 
lensed images (or arcs) coming from 9 sources. 
Each pixel contributes as two data points ($\vect{\theta_x}$ and $\vect{\theta_y}$).
The shear is computed on a $N_{\gamma}=25\times25$ grid over a field of view expanding 6 arcmin. 
Each shear measurement contributes also with two data points ($\vect{\gamma_1}$ and $\vect{\gamma_1}$)
The data vector, $\vect{\Phi}$ is then an N-dimensional vector with $N=2N_{\theta} + 2N_{\gamma} = 
2\times612 + 2\times625 = 2492$.  
The number of unknowns $N_x$ is the number of cells (or basis) $N_c$ plus two times the number of 
sources $N_s$ (the factor 2 coming from the $x$ and $y$ component), 
$N_x = N_c + 2N_s = 2\times500 + 2\times9 = 1018$ where we have assumed that 
the lens plane has been divided in 500 cells. 
The matrix $\matr{A}$ ($=\matr{\Gamma}^t\C^{-1}\matr{\Gamma}$) 
is a $N_x\times N_x$ matrix, and the vectors 
$\x$ and $\vect{a}=\G^t \C^{-1} \vect{\Phi}$ have dimension $N_x$ 
($\matr{A}$ and $\vect{a}$ are defined below equation \ref{eq_R2}).

\section{Results}  
\begin{figure}  
   \epsfysize=8.cm   
   \begin{minipage}{\epsfysize}\epsffile{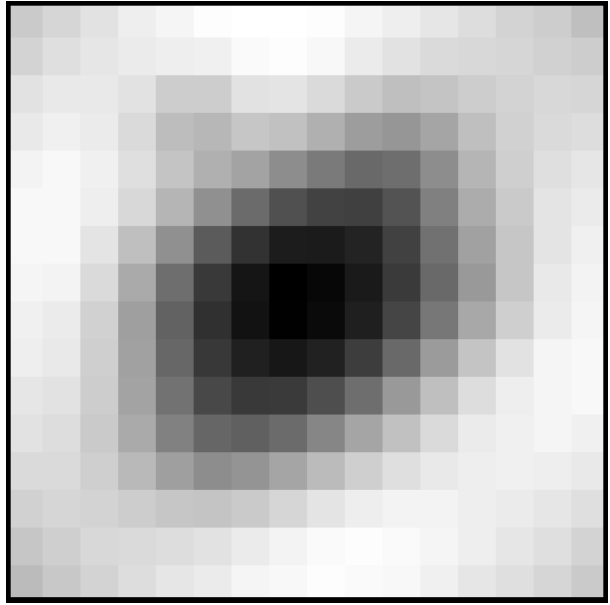}\end{minipage}  
   \caption{Recovered mass (6 arcmin) after first minimization  (regular grid). 
            The total mass is 20\% smaller than the true one. 
           }  
   \label{fig_mass0}  
\end{figure}

\begin{figure}  
   \epsfysize=8.cm   
   \begin{minipage}{\epsfysize}\epsffile{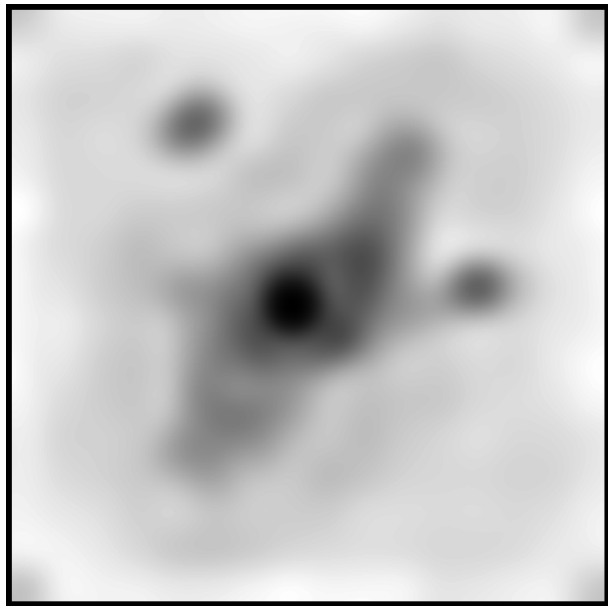}\end{minipage}  
   \caption{Recovered mass (6 arcmin) after 10 minimizations (multi-resolution grid). 
            The total mass is 2\% smaller than the true one. This result was obtained 
            using the bi-conjugate gradient algorithm (1-2
            seconds). The minimization is stopped  
            when $\chi^2 \sim 10^{-11}$. 
            Using quadratic programming, the result is very similar
            (see figure \ref{fig_massN_QADP}) but it takes several hours to converge. 
           }  
   \label{fig_massN_Biconj}  
\end{figure}

As in papers I and II, we start the minimization process assuming we know nothing 
about the mass distribution and use a regular grid to divide the lens plane. 
Also, as explained in paper I, a regular grid has the inconvenience that the small 
details of the mass distribution can not be described with enough accuracy. That means, 
the lens is less adaptable and will have problems reproducing the data. To avoid 
getting a very biased solution, the minimization process has to be stopped earlier than 
in the case where the grid reproduces finer details (bigger $\epsilon$). 
Otherwise we will end up with an unphysical solution which tries to fit
the data superposing  
big ``chunks'' of dark matter in the lens plane. Figure
\ref{fig_mass0} shows the result after the first  
iteration. It also shows the grid used to decompose the plane of the
surface mass into cells  
($N_c = 256$ cells). The first iteration finds an elliptical
distribution of mass in  
the right location but is unable to unveil any of the finer details of
the mass distribution.  
The total mass in this first iteration is smaller than the original mass by 20\%. 
Once we have a guess for the mass distribution, the adaptive grid can be 
constructed by splitting the cells with higher densities into smaller
cells. Cells are split in an iterative process which subdivides the cells having 
higher densities into four smaller sub-cells. The splitting procedure stops when the goal 
number of cells is achieved, say $N_c = 500$. 
Each time a new grid is built, the $\G$ matrix has to be recomputed again. 
Each minimization step (new grid + new $\G$ + new  
solution) usually takes about 10 seconds on a 1 GHz processor. In
figure \ref{fig_massN_Biconj}   
we show the result after 10 minimizations. The number of cells used in
this case was $N_c = 500$. Note  
how the recovered mass reproduces well most of the original structure
up to the limits of the field of  
view (compare with figure \ref{fig_mass}). 
  
\begin{figure}  
   \epsfysize=8.cm   
   \begin{minipage}{\epsfysize}\epsffile{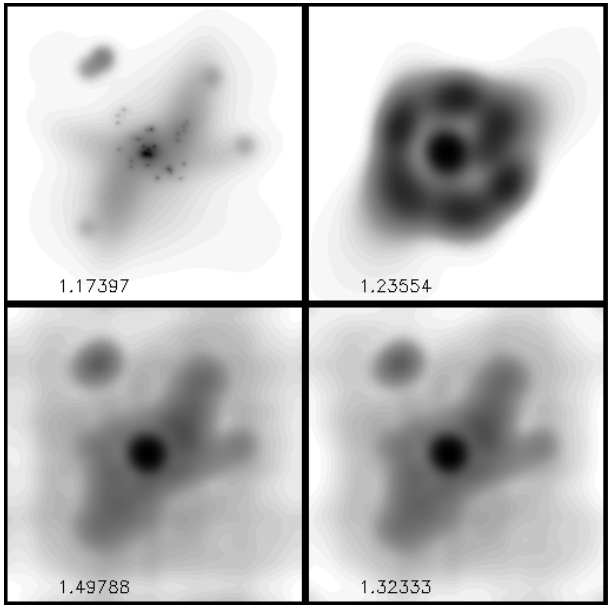}\end{minipage}  
   \caption{
            Results obtained with the QADP algorithm and using a regular grid of 
            $32\times32$ cells. The field of view is 6 arcmin in all cases. 
            From top to bottom and left to right, original mass, 
            reconstructed mass with SL data only, reconstructed mass with WL only, 
            and reconstructed mass with combined WL and SL data. The numbers at the bottom 
            of each panel is the total mass in the field of view. QADP was left running 
            until convergence (relative change in the total mass of $10^{-9}$).  
           }  
   \label{fig_massN_QADP}  
\end{figure}

Minimizing $\chi^2$ using the nonnegative quadratic programming
algorithm described above renders very similar results but the 
process can take up to several hours to converge. The main advantage of using 
QADP is that the solution converges to a mass distribution which is less biased 
with respect to the true mass than the {\it point source solution} given by the 
bi-conjugate gradient algorithm.
In figure \ref{fig_massN_QADP} we show the results obtained with QADP in the different 
scenarios, using SL data alone, using WL data alone and combining both. The combination gives 
a better reconstructed mass than the other two. Note also, how using WL data alone 
over-predicts the total mass by almost 30\% while the combination overpredicts the mass 
by only 12\%.
QADP does not suffer of the regularization problems of bi-conjugate gradient. There is 
no point source solution and the algorithm can be left running until convergence is achieved. 
The results presented above, correspond to the solution at the convergence point (relative 
change in the total mass less than $10^{-9}$). 

\begin{figure}  
   \epsfysize=8.cm   
   \begin{minipage}{\epsfysize}\epsffile{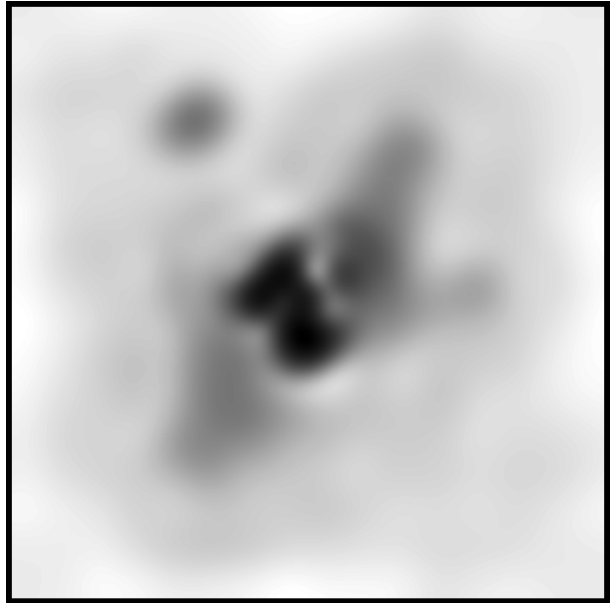}\end{minipage}  
   \caption{  
            Point source solution in the 6 arcmin field of view 
           obtained with the biconjugate gradient algorithm 
           (absolute minimum of $\chi2 \approx 10^{-13}$). 
            The total mass is only 10\% larger than the true one. 
           }  
   \label{fig_MassIterPS}  
\end{figure}

By comparing these results with the ones in papers I and II we see an
improvement in the recovered  
mass profile. First, adding weak lensing allows the reconstruction to
extend much further  
than the case where only strong lensing data is used. Second, in
papers I and II we showed  
how the results may depend on the specific choice of $\epsilon$. 
In particular, we showed that setting a very small $\epsilon$ 
produces a solution where the recovered sources are too small. This solution was called 
the {\it point source solution} in the previous papers I and II. 
Adding weak lensing partially solves this problem (see figure 
\ref{fig_MassIterPS}). The dependency with $\epsilon$ is much weaker
when weak and strong  
lensing are combined together. 
When only strong lensing is used in the minimization (see papers I and
II), the bi-conjugate gradient  
naturally tends to increase the mass in the center of the lens so the sources get more 
{\it compressed} in the center of the image (smaller $\chi^2$). Adding weak lensing 
avoids the mass to grow too much in the center since that would not
reproduce properly the  
observed shear field. On the other hand, using weak lensing alone has
the potential problem of  
the mass-sheet degeneracy. Adding strong lensing acts as a
regularizing component since a very specific amount of mass is needed
in the central region to {\it focus} the big arcs into compact sources
at different redshifts.\\ 

Another important difference with papers I and II is that they 
used no covariance matrix (or more specifically, they 
assumed that $\C = \matr{I}$). The main reason  
to introduce a covariance matrix in the present paper is to properly weight
the strong and weak lensing data.  
The covariance matrix can be also viewed as a way to allow for the instrumental 
noise and systematic error to play a role in the strong and weak
lensing data, making one data  
set more relevant than the other if their measurements are more accurate. 

\subsection{Dependence on the covariance matrix, $\C$}
\begin{figure}  
   \epsfysize=6.cm   
   \begin{minipage}{\epsfysize}\epsffile{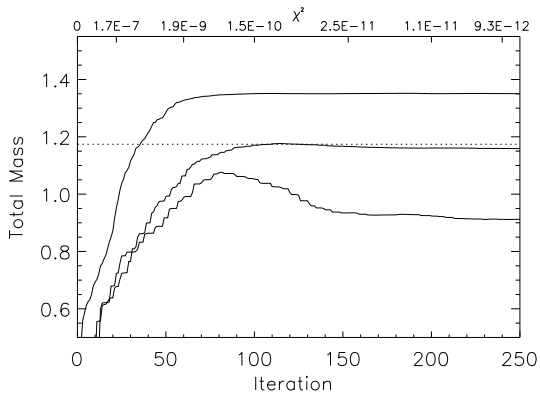}\end{minipage}  
   \caption{  
            Total mass as a function of iteration number inside the bi-conjugate 
            gradient routine. The true mass is shown as a flat dotted line. 
	    The three solid lines correspond to 3 assumptions on the covariance 
            matrix $\C$. Middle curve, $\sigma_{\theta}=3$
            pixels and $\sigma_{\gamma}=0.005$.  
            Upper curve, $\sigma_{\gamma}$ is 10 times smaller (WL
            dominates). Lower curve,  
            $\sigma_{\gamma}$ is 10 times larger (SL dominates). The
            upper x-axis shows the  
            corresponding value of the $\chi^2$ at each iteration.
           }  
   \label{fig_MassIter}  
\end{figure}

\begin{figure}  
   \epsfysize=6.cm   
   \begin{minipage}{\epsfysize}\epsffile{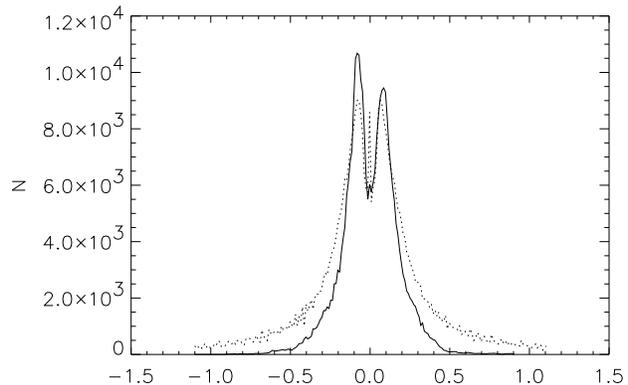}\end{minipage}  
   \caption{Histograms of the elements of $\matr{\Upsilon_y}$ (solid) and $\matr{\Delta_2}$ 
            matrices. $\matr{\Upsilon_y}$ has been multiplied by a factor 500. 
            This factor is roughly the ratio between $\sigma_{\theta}$ 
            (3 pixels or 2.1 arcsec = $10^{-5}$ rad) and  $\sigma_{\gamma}$ (0.005) 
            in the covariance matrix $\C$. These values for $\sigma_{\theta}$ and 
            $\sigma_{\gamma}$ make the $\C^{-1}\matr{\Upsilon}$
            and  $\C^{-1}\matr{\Delta}$  
            to have more or less the same variance. We refer to this case as the 
            {\it equal variance approach}.
	    The histogram of $\matr{\Upsilon_y}$ extends up to $\pm 40$. 
	    Large values in $\matr{\Upsilon}$ occur when the shear is measured 
            near the caustics. 
           }  
   \label{fig_Histo1}  
\end{figure}

The covariance matrix $\C$ controls which information is more relevant in 
the $\chi^2$. The main advantage of a combined weak+strong lensing analysis 
is that we can get both the gradient of the mass distribution up to large 
radii from the weak lensing part and the overall mass normalization plus detailed 
internal distribution from the strong lensing part. The two regimes are 
properly weighted through the covariance matrix, $\C$. 
Giving more importance to the strong lensing data will produce a better estimate in the 
central regions but will produce a result relatively insensitive to the outer regions. 
On the other hand, increasing the relevance of the weak lensing will constrain better 
the outer regions but at the expense of losing accuracy in the normalization. 
A good example of this is shown in figure \ref{fig_MassIter}. In this example we vary 
the amplitude of the covariance of the shear map by two orders of magnitude. 
Making the shear covariance smaller increases the relative importance
of the weak lensing  
in the minimization. On the other hand, increasing the shear
covariance, reduces the overall  
importance of the weak lensing part.
We found that values of the strong lensing and shear covariance around 
$\sigma_{\theta} = 10^{-5}$ rads ($\sim 2$ arcsec) and $\sigma_{\gamma}=0.005$ 
produce an unbiased estimate of the total mass (see figure \ref{fig_MassIter}). 
The ratio of these two values is equal to 
$\sigma_{\gamma}=0.005/\sigma_{\theta} = 500$. This ratio makes the 
contribution of the weak and strong lensing more or less equal in the $\chi^2$ 
(see figure \ref{fig_Histo1}). We call this the  {\it equal variance approach} since the 
dispersion of the $\C^{-1}\matr{\Upsilon}$ and  $\C^{-1}\matr{\Delta}$ is more 
or less the same (see figure \ref{fig_Histo1}). 
Also in figure \ref{fig_MassIter} we show the evolution of $\chi^2$
with the iteration number.  
The $\chi^2$ decreases quickly in the first iterations and reaches a
{\it plateau} afterward.  
The bottom of the plateau is at $\chi^2 \approx 10^{-13}$. 
This solution corresponds to the {\it point source solution}
identified in papers I and II  
(see figure \ref{fig_MassIterPS}).
In opposition to what happened in the previous papers I and II, the
point source solution  
does not deviate much from the real mass distribution. This is an important improvement  
since it shows how the combination of weak and strong lensing
stabilizes the solution and waives  
the need of any prior on the size of the sources (this prior was
needed in papers I and II).  

\begin{figure}  
   \epsfysize=6.cm   
   \begin{minipage}{\epsfysize}\epsffile{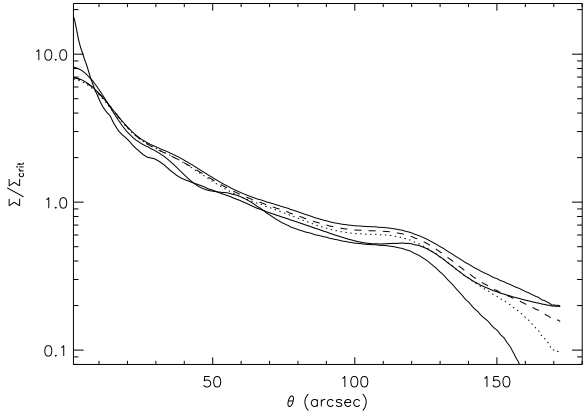}\end{minipage}  
   \caption{  
           Original profile (thick solid line) vs reconstructed ones. 
           Dotted line is the recovered profile using the fast
           bi-conjugate gradient algorithm 
           while  dashed line is the result obtained with QADP.
	   The top thin solid line corresponds to the upper curves of figure 
           \ref{fig_MassIter} (WL dominated case). The bottom thin solid line 
           correspond to the lower curve of figure 
           \ref{fig_MassIter} (SL dominated case). 
           Note how the strong lensing dominated analysis reproduces better 
           the central peak but fails in the tails and how the
           situation reverses when we increase  
           the relative importance of the weak lensing in the
           covariance matrix $\C$. 
           }  
   \label{fig_prof1}  
\end{figure}

The recovered 1-dimensional profiles show clearly the effect of
changing the relative weights of  
the shear and strong lensing data. In figure \ref{fig_prof1} we show
the same cases as in figure  
\ref{fig_MassIter}. The upper thin solid line corresponds to the case
where the weak lensing is  
given more relative weight while the lower thin solid line is the
opposite case where the strong  
lensing is given more importance. The two middle profiles (dotted and
dashed line) correspond to the  
intermediate case where the histograms of $\matr{\Upsilon}$ and
$\matr{\Delta}$ (re-scaled by $\C^{-1}$)   
share more or less the same scale (figure \ref{fig_Histo1}).

\subsection{Alternative choices for $\C$}
\label{CchoiceSec}
So far we have considered only the case where $\sigma_{\theta}$ and $\sigma_{\gamma}$ 
are constants. The {\it equal variance approach} consists on choosing $\sigma_{\gamma}$ 
such that the dispersion of the $\C^{-1}\matr{\Upsilon}$ and
$\C^{-1}\matr{\Delta}$  
matrices are more or less the same. In other words, this choice for $\C^{-1}$ 
makes the contribution from the weak and strong lensing more or less equal 
(when $N_{\theta} \approx N_{\gamma}$). 
This choice produces satisfactory results as we have seen above.\\
Since $\C$ can be seen as the matrix containing the covariances
of the data points,  
one may feel tempted to play with different weights for the data
set. For instance, one may  
consider giving more relative importance to the smaller radial arcs than to the bigger 
tangential arcs. This is motivated by the fact the the residual of the
strong lensing part  
is more clearly dominated by the big tangential arcs than by the small 
radial ones. We have tried different weighting factors in the matrix
$\C$ and found that  
the best results are obtained when the weight of the strong lensing
data, $\sigma_{\theta}$,  
is homogeneous over the field of view , that is, 
all data points are given the same importance independently or whether
they are forming part of a  
giant arc or a tiny radial arc. Weighting the radial arcs more than
the tangential ones produces  
biased results in the recovered mass distribution, included the
position of the central peak.  
A good result is obtained also when the weight is proportional to the
fraction of pixels in  
the system compared with the total number of pixels in all systems. In
this case, the results  
are very similar to the ones obtained with an homogeneous weight in $\C$.


\subsection{Dependence on the basis $f_l$}
\begin{figure}  
   \epsfysize=6.cm   
   \begin{minipage}{\epsfysize}\epsffile{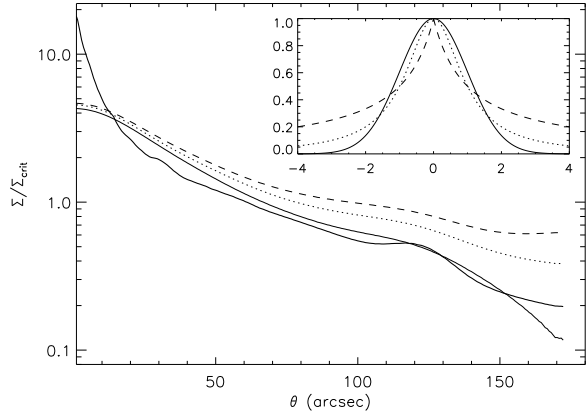}\end{minipage}  
   \caption{  
           Original profile (thick solid line) vs reconstructed ones. 
	   From top to bottom. Using as basis $f_l$ isothermal spheres (dashed), 
           power laws ($r^{-2}$) (dotted) and Gaussians (thin solid line).
           The inner plot shows the three basis for the same scale $\sigma = 1$. 
           Gaussian (solid), power law $(r^2 + \sigma)^{-1}$ 
           (dotted) and isothermal $(r + \sigma)^{-1}$ (dashed). Basis
           with extended tails  
           act adding a constant surface mass density to the overall mass. Compact 
           functions like the Gaussian can concentrate the mass closer to the cell where 
           they are positioned.
           }  
   \label{fig_prof2}  
\end{figure}

In this section, we will discuss the role of the basis functions $f_l$ used to decompose 
the mass (equation \ref{eq_F}).\\ 
We found that in general compact basis give better results than extended ones. 
As an example, in figure \ref{fig_prof2} we show the reconstructed profiles using three 
different sets of basis functions: \\
i) A Gaussian basis centered in each cell with a width, $\sigma$, equal to two times 
the size of the cell, 
\begin{equation}
G(r) \propto {\mathrm exp}(-r^2/2\sigma^2).
\end{equation}
ii) An isothermal sphere with a core of the same scale  $\sigma$,
\begin{equation}
I(r) \propto \frac{1}{r + \sigma}.
\end{equation}
iii) A power law also with a core of the same scale $\sigma$,
\begin{equation}
P(r) \propto \frac{1}{r^2 + \sigma}.
\end{equation}
The results obtained with the isothermal sphere and the power law show
a constant sheet excess  
in the surface mass density which is probably due to the extended
tails of the basis. These  
two basis reproduce well the central parts but fail in predicting the
right density in the outer  
regions. This behavior may be a manifestation of the
mass-sheet degeneracy.  

\section{Conclusions}  

In this paper, we have presented a way of consistently combining strong and weak lensing 
using a non-parametric method (WSLAP) which does not rely on any prior on the luminosity 
and does not suffer from significant regularization problems. Finding the solution through 
the bi-conjugate gradient still is affected by minor regularization problems as the 
minimization has to be stopped before the absolute minimum is reached. This is needed to 
avoid the {\it point source solution}. However, we have seen how even the point source 
solution can be a good estimation of the mass when weak and strong lensing are combined. 
In previous papers using only strong lensing, we found that the point source solution obtained 
with the bi-conjugate gradient was a bad estimate of the mass. 
On the other hand, the solution obtained with QADP does not suffer of regularization problems. 
Imposing the constraint that the masses have to be positive is a natural way to regularize 
the solution. \\

Adding weak lensing has two major effects on the solution; i) when minimizing 
the quadratic function with standard algorithms (for instance the bi-conjugate gradient) 
the result is basically insensitive to the threshold $\epsilon$ where
the minimization is stopped  
since the negative masses which appear when $\epsilon$ is too small
can not reproduce the shear  
field properly, 
ii) the profile can be better reproduced inside and beyond the position of the 
big arcs. The weak lensing data allow us to eliminate the use of any {\it prior} on the 
physical size of the sources and to better constrain the range of solutions, 
thus adding more robustness to the final result.\\

The method allows the freedom to make two choices, for the covariance
matrix $\C$ and  the basis functions $f_l$, and we quantified the impact 
of both on systematic errors in the mass reconstruction. 
We found that the {\it equal variance approach} for the covariance  
matrix renders satisfactory results. Giving more relative importance
to the radial than to  
the tangential arcs produces a biased solution for the mass. Weighting
the arc systems proportional to  
their area in the sky produces similar results as in the case of the
equal variance approach.  
Regarding the basis functions, we found that functions $f_l$ which
are compact produce better  
results than extended functions, specially in describing the
weak lensing part of the data.  
This fact may be a manifestation of the mass-sheet degeneracy in
the weak lensing data.\\ 

This paper is more of an illustration of how to extend the methodology of SLAP 
(papers I and II) to include weak lensing than a detailed description
of the capabilities  
and failures of the WSLAP approach. However, although an illustration,
this paper demonstrates  
the usefulness of non-parametric methods when combining weak and strong lensing.

Much work needs still to be done to address possible 
systematic issues, but as described in paper II, most of this work will
have to be done when  
WSLAP is applied to real data. The systematics may depend on the
specific nature of the problem  
(number of sources, geometry and redshift of the lens, quality of the data). 
Future improvements will include adding photometric information and a 
better modeling of the sources (Sandvik et al. in preparation). 

WSLAP is now available to the community at: \\

\ \ \ \ \ \ \ http://darwin.cfa.harvard.edu/SLAP/. 

\section{Acknowledgments}  
This work was supported by NSF CAREER grant AST-0134999, NASA grant  
NAG5-11099, the David and Lucile Packard Foundation and   
the Cottrell Foundation. The authors would like to thank David Hogg for 
useful discussions on quadratic programming and Elizabeth E. Brait for 
useful comments.
  


\bsp  
\label{lastpage}  
\end{document}